\documentclass{article}
\usepackage{graphicx} 
\usepackage{amsmath, amsthm, amssymb}
\usepackage{diagbox}
\usepackage{caption}
\usepackage{tikz}
\usepackage{multicol}
\usepackage[style=numeric]{biblatex}
\newtheorem{lemma}{Lemma}
\newtheorem{theorem}{Theorem}
\newtheorem{corollary}{Corollary}

\usepackage[a4paper]{geometry}
\newcommand{\lem}[2]{\begin{lemma} \label{#1} #2 \end{lemma}}
\newcommand{\thm}[2]{\begin{theorem} \label{#1} #2 \end{theorem}}
\newcommand{\crl}[2]{\begin{corollary} \label{#1} #2 \end{corollary}}

\newcommand{\refl}[1]{Lemma \ref{#1}}

\newcommand{\reftb}[1]{Table \ref{#1}}
\title{Cut a Numeric String into Required Pieces}
\author{CAI Yinqi\\
School of Software Engineering \\
Southeast University\\
Nanjing, China \\
Email: yinqicai02@gmail.com\\}
\date{\today}
\begin{document}
\maketitle
\begin{abstract}
We study the problem of cutting a length-$n$ string of positive real numbers into $k$ pieces so that every piece has sum at least $b$. The problem can also be phrased as transforming such a string into a new one by merging adjacent numbers.
We discuss connections with other problems and present several algorithms in connection with the problem:
an $O(n)$-time greedy algorithm, an $O(kn\log n)$-time dynamic programming algorithm, and an $O(n+ k \log n + 2^kk)$-time FPT algorithm for $n-k$ pieces.
\end{abstract}

\section{Introduction}
Let us begin with a line segment with 10 marked points in Figure~\ref{fig:line}
(you may regard the line segment as a piece of stick or rope)
and consider the following problems:

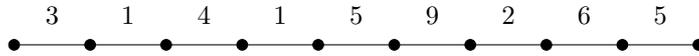
\begin{figure}[ht]
  \centering
  \begin{tikzpicture}
    \draw (0,0) -- (9,0);
    \foreach \x/\label in {0/3, 1/1, 2/4, 3/1, 4/5, 5/9, 6/2, 7/6, 8/5, 9/}
    {
      \filldraw (\x,0) circle (2pt);
      \ifnum\x<9
        \node[above] at (\x+0.5,0.15) {\label};
      \fi
    }
  \end{tikzpicture}
  \caption{Line segment with marked points, where a number between two adjacent points
  indicates their distance.}{}\label{fig:line}
\end{figure}

\begin{enumerate}
    \item Use locations indicated by marked points only to cut the line segment into 3 pieces as evenly as possible.
    \item Select 4 marked points from the line segment to maximize the smallest pairwise distance of the selected points.
    \item Remove as few marked points as possible from the line segment so that remaining marked points are at least distance 8 apart.
\end{enumerate}

As we will see in Section 2, the above three problems are interconnected and can be formulated as problems on a sequence of positive real numbers, referred to as a {\em string}, with respect to the following operation of merging adjacent numbers in the string: replace two adjacent numbers in the string by their sum.
For example, the line segment of Figure~\ref{fig:line} can be represented by string 
$S = 3, 1, 4, 1, 5, 9, 2, 6, 5$.
By performing 6 merges, we obtain a new string $14, 11, 11$ from $S$, 
which yields a solution of problem 1.

We are interested in performing a small number of merges to raise the value of the smallest number in the string as much as possible,
which brings us the following two related problems we study in the paper:

\noindent {\bf Fewest-Merge}($b$): Find fewest merges for a string so that every number in the resulting string is at least $b$.

\noindent {\bf MaxMin-Merge}($k$): Find $k$ merges in a string to maximize the minimum number in the resulting string.

For our example string $S$, a solution to {\bf Fewest-Merge}(5) consists of 4 merges to produce a new string 8, 6, 9, 8, 5, 
and a solution  to {\bf MaxMin-Merge}(3) produces a string 4, 5, 5, 9, 8, 5 with the best possible minimum value 4.
 
\noindent {\bf Related work}:
Our merge problems have close connections with several other problems: the well-studied {\sc MaxMin Diversity} problem\footnote{The {\sc MaxMin Diversity} problem aims at finding $k$ points from a set of $n$ points to maximize the minimum pairwise distance of the selected points.} for a set of points on a line, edge contraction of vertex-weighted paths, and cutting a string into substrings.
And the three problems in the beginning of the paper can be formulated as {\bf Fewest-Merge} and {\bf MaxMin-Merge} problems.

{\sc MaxMin Diversity} has been studied extensively~\cite{Mart, Marti, Poru, Kath} mainly in the operations research community, and it has a wide range of applications.
Most work for the problem is on approximation and experimental algorithms, and we do not find any work for the problem for points on a line.

We find some webpages mentioning edge contractions in vertex-weighted graphs and several merge operations, but, surprisingly, are unable to find any research paper 
concerning our merge problems for numeric strings.

\noindent {\bf Main results}: We give a linear-time algorithm for {\bf Fewest-Merge}($b$) by considering a string partition problem, and use dynamic programming to obtain an $O((n-k)n\log n)$-time algorithm for {\bf MaxMin-Merge}$(k)$ by converting it to {\sc MaxMin Diversity} 
for a set of points on a line.
We also present an $O(n+ k \log n + 2^kk)$-time algorithm
for {\bf MaxMin-Merge}$(k)$, which is linear time for each fixed value of $k$ and faster than our dynamic programming algorithm for small values of $k$.

\noindent {\bf Organization}: We discuss connections with other problems in  Section 2, consider {\bf Fewest-Merge}($b$) and {\bf MaxMin-Merge}$(k)$, respectively, in Sections 3 and 4, and propose a few open problems in Section 5 to conclude the paper.

\noindent {\bf Definitions}: A few definitions are in order to prepare for our technical discussions in the following sections.
A string $S$ in this paper consists of a sequence of positive real numbers.
We use $m(S)$ to denote the minimum number in $S$, and $\Sigma(S)$ the sum of all numbers in $S$.
A substring of $S$ is a {\em $b$-string} if its sum is at least $b$,
and a {\em $b$-partition} of $S$ cuts $S$ into $b$-strings.
A {\em $b$-prefix} $s$ of $S$ is a prefix of $S$ that forms a $b$-string, and $s$ is
the {\em minimal $b$-prefix} if the deletion of the last number from $s$ reduces its sum to less than $b$.

\section{Connections with other problems}
Our problems {\bf Fewest-Merge}($b$) and {\bf MaxMin-Merge}($k$) have close connections to other problems: partition of a string into substrings,
selection of points to maximize minimum pairwise distance, and edge contraction in vertex-weighted graphs.
We now discuss these connections which enables us to see interconnections among the 3 problems in the beginning of the paper, and are also vital for obtaining our algorithms 
in the paper.

First we note that partition and merge are two different ways to describe cutting a string into pieces.
Recall that a {\em $b$-partition} of $S$ cuts $S$ into $b$-strings, i.e.,
substrings with sum at least $b$.

\lem{cut-merge}{
A string $S$ has a $b$-partition $\Pi$ into $t$ $b$-strings iff it can be transformed into a $b$-string $S^*$ by $|S| - t$ merges.
}
\proof{The lemma directly follows from the following one-to-one correspondence between $b$-strings in $\Pi$ and numbers in $S'$:
For a $b$-string $S'$ in $\Pi$, it takes $|S'| - 1$ merges to merge them into a single number $s'$ of $S^*$.
\qed
}

Next we can represent a string $S = s_1, \dots, s_n$ by $n+1$ points 
$P(S) = p_0, \dots, p_n$ on a line,
where $p_0 = 0$ and $p_i = \sum_{j=1}^i s_j$ for each $1 \le i \le n$.
This associated point set $P(S)$ of $S$ can be used as cut points to cut $S$ into substrings, which enables us to
tie partition of $S$ with the following well-known problem for points on a line:
\begin{quote}
    {\sc MaxMin Diversity}\\
    {\bf Instance}: Set $P$ of $n$ points.\\ 
    {\bf Task}: Select $k$ points $P'$ from $P$ to maximize the minimum pairwise distance of $P'$. \\
\end{quote}

\lem{cut-diversity}{
A string $S$ admits a partition into $k$ substrings each has sum at least $d$ iff $P(S)$ has $k+1$ points where every pairwise distance is at least $d$.
}
\proof{By the definition of $P(S)$, $P(S) -\{p_0, p_n\}$ is exactly the set of possible
positions to cut string $S$ into substrings in terms of distances of these positions from the beginning of the string.
This gives us a one-to-one correspondence between partitions 
$[S_1, \dots, S_k]$
of $S$ into $k$
pieces and $k-1$ points $P'$ from $P(S) - \{p_0, p_n\}$.
Since sums of substrings in a partition correspond to distances between 
adjacent points in $P' \cup \{p_0, p_n\}$, and the minimum pairwise distance
for any set of points on a line equals the minimum distance between adjacent points,
the equivalence in the lemma follows.
\qed}

As for a connection with edge contraction in vertex-weighted graphs, we can take each
number $s_i$ in string $S$ as a vertex $v_i$ with weight $s_i$, and add edge 
$v_iv_{i+1}$ for each $1 \le i \le n-1$ to form a path.
Then merging two adjacent numbers $s_i$ and $s_{i+1}$ in $S$ is the same as contracting edge $v_iv_{i+1}$ in the path to form a new vertex with weight $s_i + s_j$.
And our merge problems for a string is just the following edge contraction problem, which  is intractable even for complete graphs, for a vertex-weighted path.

\begin{quote}
{\sc Merging Vertices} \\
{\bf Instance}: Weighted graph $G = (V,E;w)$ with $w: V \rightarrow R^+$, positive real number $b$ and positive integer $k$. \\
{\bf Question}: Can we perform at most $k$ edge contractions in $G$ to obtain a new graph where every vertex has weight at least $b$?
\end{quote}

\thm{npc}{The {\sc Merging Vertices} problem is NP-complete for weighted complete graphs
}
\proof{
We give a reduction from the classical {\sc Partition} problem.
For an arbitrary instance $S = \{s_1, \dots, s_n\}$ of {\sc Partition}, we construct a weighted complete graph $G$ on $n$ vertices $\{v_1, \dots, v_n\}$ with $w(v_i) = s_i$,
set $k = n - 2$ and $b = \frac{1}{2}\sum_{i=1}^n s_i$.

Suppose that $\{S', S-S'\}$ is a valid partition for {\sc Partition}.
We can merge all vertices corresponding to $S'$ into one vertex, and vertices
corresponding to $S - S'$ into another vertex.
These two merged vertices both have weight $b$, and the number of edge 
contractions is $n-2$, implying that $G$ is a yes-instance of {\sc Merging Vertices}.

Conversely suppose that we can use $n-2$ edge contractions to obtain a graph on two
vertices $v^1$ and $v^2$ both of weight at least $b$.
Let $S'$ be numbers in $S$ corresponding to vertices merged into $v^1$.
Then $\{S', S-S'\}$ forms a valid partition of $S$. \qed
}

\section{Cut into $b$-strings}

In this section we study {\bf Fewest-Merge}($b$) of transforming a string $S$ into a $b$-string by fewest merges, where 
a {\em $b$-string} is a string with every number at least $b$.
We will present a linear-time algorithm for the problem by considering
an equivalent problem of cutting a string into $b$-strings.

We first observe the following two extreme cases of the problem that relates the number $k$ of merges with the number $n_b$ of integers in $S$ smaller than $b$:
\begin{description}
\item [$k=n_b$:] $S$ has $n_b$ numbers less than $b$, and none of them are adjacent. Each such number is merged with an adjacent number, and we need $k$ merges.
\item [$k =n_b/2$:] $S$ has $n_b$ numbers less than $b$ which can be paired up and each pair has sum at least $b$. We can merge each pair
to obtain a number at least $b$, and the number of merges is $n_b/2$.
\end{description}
Since every number less than $b$ requires one merge, the above observations bound the number of merges between $n_b/2$ and $n_b$.

By \refl{cut-merge}, {\bf Fewest-Merge}($b$) is equivalent to the problem
of partitioning $S$ into as many $b$-strings as possible, and we will focus on this partition problem.
Recall that the {\em minimal $b$-prefix} of $S$ is a prefix $s$ of $S$ with sum at least $b$ but the deletion of the last number from $s$ reduces its sum to less than $b$.
The following two results reveal the importance of minimal $b$-prefix in obtaining a linear-time algorithm for our partition problem.

\lem{morethan2}{
A string $S$ has a $b$-partition into at least two $b$-strings iff for the minimal $b$-prefix $s$ of $S$, $S - s$ has sum at least $b$.
}
\proof{The condition is obviously sufficient as $[s, S - s]$ is a valid $b$-partition.
Conversely, if $S$ has a $b$-partition $[S_1, S_2, \dots ]$, then $S - S_1$ has sum at least $b$,
implying that $S - s$ has sum at least $b$ as $s$ is a substring of $S_1$. \qed
}

A $b$-partition is an {\em optimal $b$-partition} when the number of $b$-strings in the partition is maximized,
and we note a useful property of an optimal $b$-partition.

\lem{opt-soln}{
 A string $S$ always has an optimal $b$-partition with the minimal $b$-prefix of $S$ as the first substring whenever $S$ admits an optimal $b$-partition with at least two $b$-strings.
}

\proof{Let $\Pi = [S_1, S_2, \dots, S_t]$ be an optimal $b$-partition of $S$,
and $s$ the minimal $b$-prefix of $S$.
Then we can replace $S_1$ and $S_2$ in $\Pi$ by $s$ and $(S_1 - s) \cdot S_2$ 
to obtain another optimal $b$-partition of $S$, which has the required property. \qed
}

The above result suggests that we can repeatedly cut off the minimal $b$-prefix in the current string to 
obtain an optimal $b$-partition, which results in the following greedy algorithm for our partition problem.

\begin{tabbing}
{\bf Algorithm} Optimal-Partition$(S, b)$ \\
{\bf Input:} String $S$ of positive real numbers, and positive real number $b$.\\
{\bf Output:} An optimal $b$-partition $\Pi$ of $S$. \\ \\

Set $\Pi$ to empty initially; \\
{\bf If} sum of $S$ is less than $b$ {\bf then print} ``no solution'' and {\bf halt}; \\
{\bf while } \= sum of $S$ is at least $b$ {\bf do} \\
\> Find the minimal $b$-prefix $s$ of $S$; \\
\> Add $s$ to the rear of $\Pi$;\\
\> $S \leftarrow S - s$\\
{\bf end while}; \\
Replace the last string $z$ in $\Pi$ by $z \cdot S$; \\
{\bf print} $\Pi$.\\
\end{tabbing}

For $S = 5, 4,  5, 7, 1, 1, 4, 4, 5, 6$ and $b =10$, the above algorithm produces an optimal 10-partition $5, 4, 5 | \; 7, 1, 1 ,4 | \; 4, 5, 6$.
If we run the algorithm on $S$ from right to left (i.e., on the reversed string of $S$), we obtain  a different optimal 10-partition: $5, 4, 5, 7 \; |1, 1, 4, 4 \;|5, 6$.
Intuitively, the latter one may have size different from the former one.
However these two $b$-partitions have the same size as guaranteed by the correctness of the algorithm.

\thm{agl-cut}{
Algorithm {\rm Optimal-Partition} correctly finds an optimal $b$-partition of a string $S$ of length $n$ in $O(n)$ time.
}
\proof{The algorithm clearly produces a $b$-partition $\Pi$, when $S$ has one,
as $\Pi$ cuts $S$ into substrings each with sum at least $b$. 
We now use induction on the length of $S$ to show that $\Pi$ is an optimal $b$-partition.
For the base case, $S$ has a single number and our algorithm works
correctly as $S$ admits a $b$-partition (just $S$ itself) iff the number is at least $b$.
Now we assume that the algorithm produces an optimal $b$-partition when $|S| < n$,
and consider a string of length $n$.

Our algorithm finds a minimal $b$-prefix of $S$, and reduces $S$ to $S - s$.
If $S -s$ has no $b$-partition then, by \refl{morethan2}, $S -s$ has sum less than $b$.
In this case, the algorithm replaces $s$ in $\Pi$ by $S$ to form the correct optimal $b$-partition $[S]$.
Otherwise, by the induction hypothesis, our algorithm finds an optimal $b$-partition $\Pi'$
of $S - s$. We show that  $\Pi = s \cup \Pi'$ is an optimal $b$-partition of $S$.

By \refl{opt-soln}, $S$ admits an optimal $b$-partition $\Pi^*$ with $s$ being the first string. Note that $\Pi^* -s$ is a $b$-partition of $S-s$, and hence has size no more than $\Pi'$ as the latter is an optimal $b$-partition of $S-s$.
It follows that $\Pi = s \cup \Pi'$ is no smaller than $\Pi^*$, and hence is indeed
an optimal $b$-partition of $S$. 
This shows the correctness of the algorithm.

For the running time, we note that our algorithm does a left-to-right scan of
the input string $S$, and we can easily implement it in $O(n)$ time by, say, using
an array for $S$ and a stack for $\Pi$. \qed
}

By \refl{cut-merge}, the above theorem directly gives us the following result 
in terms of the number of merges, i.e., a linear-time algorithm for {\bf Fewest-Merge}($b$).

\crl{merge}{It takes $O(n)$ time to find fewest merges to transform
a string of length $n$ into a $b$-string.
}

\section{Cut into $k$ pieces}

We now consider {\bf MaxMin-Merge}$(k)$ and
use dynamic programming to obtain an $O((n-k)n\log n)$-time algorithm by converting the problem to the {\sc MaxMin Diversity} problem points on a line.
We also present an $O(n+ k \log n + 2^kk)$-time algorithm, which is linear time for each fixed value of $k$ and faster than our dynamic programming algorithm for small values of $k$.

\noindent {\bf Remark}. If number $b$ and all numbers in the string $S$ are integers,
we can use our linear-time algorithm for {\bf Fewest-Merge}$(b)$ as a subroutine to solve {\bf MaxMin-Merge}$(k)$
by using a binary search on value $b$, which results in 
an $O(n \log b)$ time algorithm for the latter problem.
One may argue that this is not a polynomial algorithm as $b$ can be very large, say $2^{2^n}$, comparing with $n$.
Nevertheless, it is unreasonable to allow numbers larger than $2^{O(n^d)}$,
where $d$ is a constant, when we use uniform cost to analyze the running time.
A reasonable assumption is that all numbers in $S$ are bounded by a polynomial of $n$,
i.e., every number in $S$ is at most $O(n^d)$ for some constant $d$.
Then the binary search approach gives an $O(n\log n)$ time algorithm.
If we use logarithmic cost, then the input size is between $n + \log b$ and 
$n\log b$ and the algorithm runs in polynomial time  with respect to this input length. 
However this binary search method does not work for real number $b$ and general $S$ consisting of real numbers.

As discussed in Section 2, partition is an alternative way to view string transformation by merging: each merged number in the transformed string corresponds to a substring in a partition.
It follows that {\bf MaxMin-Merge($(n-k)$)} is equivalent to 
the following partition problem:

{\sc Cut String[$k$]}: Cut a string $S$ into $k$ substrings to maximize the minimum sum of the $k$ substrings.

For the above problem, as we have seen in Section 2,
it is convenient to also use the associated point set 
$P(S) = p_0, \dots, p_n$ of $S$,
where $p_0 = 0$ and $p_i = \sum_{i=1}^n s_i$ 
for each $1 \le i \le n$.
Points in $P(S)$, except endpoints $p_0$ and $p_n$, 
are candidate cut points for $S$: any $k$ points from $P(S)$
form a {\em $k$-cut} that can be used to cut $S$ into $k+1$ substrings.
Therefore  {\sc Cut String[$k$]} is equivalent to the problem of 
finding $k-1$ points in $P(S)$ to cut the line segment from point 0 to point $p_n$
into $k$ pieces to maximize the the length of the shortest piece.

Before we present an $O(kn\log n)$-time algorithm for {\sc Cut String[$k$]}, we describe linear algorithms for $k = 2, 3$
to give us a better understanding of the problem.

To solve {\sc Cut string[2]}, we need to find a point $p_i$ from $P(S)$ to make $\min\{p_i, p_n - p_i\}$ as large as possible.
Let $\delta$ be the distance between $p_i$ and 
the middle point $p_n/2$ of $P(S)$.
Then the smaller one of $p_i$ and  $p_n - p_i$ equals $p_n/2 - \delta$, 
which is maximized when $\delta$ is minimized.
Therefore $p_i$ is a point closest to the middle point, and we can easily find such a point in $O(n)$ time.

We can also solve {\sc Cut String[3]}
in $O(n)$ time, instead of the trivial $O(n^2)$ time algorithm by considering  $O(n^2)$ possible cuts into 3 pieces.
Note that $P(S)$ has two trisection points $\frac{1}{3}p_n$ and $\frac{2}{3}p_n$.
It turns out that points in $P(S)$ that are adjacent to trisection points
play an important role in an optimal partition.

\lem{near-trisection}{
A string always admits an optimal $2$-cut
where one cut point is either equal or adjacent to a trisection point.
}
\proof{Keep in mind that any 2-cut results in a piece of length at most $p_n/3$.
Let $p,q$, where $p < q$, be the two cut points of an optimal 2-cut
and $t_1,t_2$, where $t_1 < t_2$, be the two trisection points of $P(S)$.
If $p \ge  t_1$ we can replace $p$ by $t_1$ if $t_1 \in P(S)$ or
the point right-adjacent to $t_1$ to obtain another optimal 2-cut
(note that the length of the first piece is reduced but still at least $p_n/3$).
Similarly we can do so for cut point $q$ when $q \le t_2$.
For the remaining case $p < t_1$ and $q > t_2$, 
the middle piece is more than $p_n/3$
and we can replace $p$ by $t_1$ if $t_1 \in P(S)$ or
the point left-adjacent to $t_1$ to obtain another optimal 2-cut
(note that the middle piece still has length at least $p_n/3$).
\qed
}

Since for each of the two trisection point, either the point (if it is in $P(S)$) or at least 
one of its two adjacent points belongs to an optimal 2-cut, we
need only try at most 4 points as a cut point and then use the linear algorithm for 1-cut to find the other cut point.
This gives us a linear algorithm for finding an optimal 2-cut of a string.

We now consider {\sc Cut String[$k$]} in general.
By \refl{cut-diversity}, it suffices to solve the following special case of {\sc MaxMin Diversity}:
Choose $k$ points $P'$ from $n$ points $P$ on a line to maximize the distance between adjacent points in  $P'$.
We use dynamic programming to solve it in $O(kn\log n)$ time.

The {\em minimum pairwise distance} of $P$, denoted mpd$(P)$, is the minimum pairwise distance in $P$,
which equals the minimum distance between adjacent points in $P$.
Let $P_i$ be the first $i$ points of $P$.
A $k$-subset of $P$ is an {\em optimal $k$-subset} if the minimum pairwise distance
in the $k$-subset is the largest among all $k$-subsets of $P$.
Define $d(i,j)$ to be mpd$(P')$ of an optimal $j$-subset $P'$ of $P_i$ that 
uses point $p_i$, and we have the following recurrence for $d(i,j)$.

\lem{recurrence}{
For any $2 \le i \le n$ and $2 \le j \le k$, we have 
\[ d(i,j) = \max_{j-1 \le i' \le i-1} \{ \min\{d(i',j-1), p_{i-1} - p_{i'-1}\}\}. \]
}
\proof{
Let $P'$ be an optimal $j$-subset of $P_i$  that uses point $p_i$, and $p_{i'}$ the point in $P'$ right before point $p_i$.
Then $p_{i'}$ is a point from interval $[p_{j-1},p_{i-1}]$ and 
\[ {\rm mpd}(P')= \min\{ {\rm mpd}(P' - p_i), p_i - p_{i'}\} \]
Since $P'-p_i$  contains point $p_{i'}$, we have $d(i',j-1) \ge {\rm mpd}(P'- p_i)$.
Now we consider two cases:

Case 1: ${\rm mpd} (P' - p_i) \ge p_i-p_{i'}$: We have ${\rm mpd}(P)=p_i-p_{i'}$, and
$\min\{d(i',j-1),p_i-p_{i'}\} = p_i-p_{i'}$, implying ${\rm mpd}(P') = \min\{d(i',j-1),p_i-p_{i'}\}$.

Case 2. ${\rm mpd}(P'-p_i) < p_i-p_{i'}$: Since $P'$ is an optimal $j$-subset of $P_i$ that contains 
point $p_j$, $P'-p_i$ is an optimal (j-1)-subset containing point $p_{i'}$,
implying ${\rm mpd}(P'-p_i) = d(i',j-1)$. Hence we also have 
${\rm mpd}(P') = \min\{d(i',j-1), p_i - p_{i'}\}$.

Since any point from interval $[p_{j-1},p_{i-1}]$ is a possible candidate for $p_{i'}$, the largest value of
$\min\{d(i',j-1),p_i-p_{i'}\}$  for all $i'$  from $j-1$ to $i-1$ equals the value of mpd$(P')$ 
and hence we have the formula in the lemma. \qed
}

With the recurrence in \refl{recurrence} in hand, we can easily
use an $n$ by $k$ array for $d(i,j)$ and fill the array row by row.
For initialization we note that $d(i,2) = p_{i-1}$ for $2 \le i \le n$.
To find an optimal $j$-solution, we do a standard bookkeeping in computing $d(i,j)$: use entry $s(i,j)$ to record the index $i'$ for achieving an optimal $j$-solution.

As an example, \reftb{dp} shows values of $d(i,j)$ for choosing 4 points from $P$ = 0, 3, 4, 6, 7, 9, 10, 13, 16.
\reftb{dp2} gives optimal indices $s(i,j)$ in bookkeeping, and 
\reftb{dp3} illustrates the process of finding an optimal solution $P'$ by using $s(i,j)$.

\begin{table}[ht]
\centering
\caption{Values of $d(i,j)$.}\label{dp}
\begin{tabular}{|c|c|c|c|c|c|c|c|c|c|}
\hline
\diagbox{$j$}{$d(i,j)$}{$i$} & 2 & 3 & 4 & 5 & 6 & 7 & 8 & 9 \\
\hline
2 & 3 & 4 & 6 & 7 & 9 & 10 & 13 & 16 \\
\hline
3 & / & 1 & 3 & 3 & 4 & 4 & 6 & 7 \\
\hline
4 & / & / & 1 & 1 & 3 & 3 & 4 & 4 \\
\hline
\end{tabular}
\end{table}

\begin{table}[ht]
\centering
\caption{Values of $s(i,j)$.}\label{dp2}
\begin{tabular}{|c|c|c|c|c|c|c|c|c|c|}
\hline
\diagbox{$j$}{$s(i,j)$}{$i$} & 2 & 3 & 4 & 5 & 6 & 7 & 8 & 9 \\
\hline
2 & 1 & 1 & 1 & 1 & 1 & 1 & 1 & 1 \\
\hline
3 & / & 2 & 2 & 3 & 3 & 4 & 5 & 6 \\
\hline
4 & / & / & 3 & 4 & 4 & 5 & 6 & 6 \\
\hline
\end{tabular}
\end{table}

\begin{table}[ht]
\centering
\caption{Computing an optimal solution from $s(i,j)$.}\label{dp3}
\begin{tabular}{|c|c|c|}
\hline
$s(i,j)$ & $P'$ \\
\hline
$s(9,4)$ = 6 & \{9\} \\
\hline
$s(6,3)$ = 3 & \{4, 9\} \\
\hline
$s(3,2)$ = 1 & \{0, 4, 9\} \\
\hline
/ & \{0, 4, 9, 16\} \\
\hline
\end{tabular}
\end{table}


It takes $O(n)$ time to compute $d(i,j)$ by using the recurrence of \refl{recurrence}, and we can reduce it to $O(\log n)$ time by observing the following property of $d(i,j)$.

\lem{monotone}{
For any integers $j-1 \le i' < i$, $d(i',j) \le d(i,j)$.
}
\proof{When $i' < i$, every $j$-subset $P'$ of $P_{i'}$ is a $j$-subset of $P_i$ and we can replace point $p_{i'}$ in $P'$ by
point $p_i$ to obtain a new $j$-subset $P^*$ for $P_i$.
Clearly $P^*$ does not reduce pairwise distance of $P'$ and the inequality follows. \qed
}

Since $p_i-p_{i'}$ is a decreasing function of $i'$ and $d(i',j)$ an nondecreasing function of $i'$,
the minimum of these two function is attained at their intersection and the above 
lemma allows us to use binary search on $i'$,
which ranges from $j-1$ to $i-1 \le n$, to find their intersection: if $d(i',j-1) > p_i-p_{i'}$ then we increase the right side of the inequality by reducing $i'$ to $i'/2$, and if
$d(i',j-1) < p_i-p_{i'}$ then we decrease the right side by moving $i'$ up to $(i'+i)/2$.
Once we have found the intersection of these two function, the optimal choice of $i'$ is between the two points
that sandwich the intersection point.

We can now state the following result by summarizing above discussions.

\thm{dp-alg}{
Both {\sc MaxMin Diversity}$(k)$ for a set of $n$ points on a line and {\sc Cut String}$[k]$ can be solved in $O(kn\log n)$ time, and {\bf MaxMin-Merge}$(k)$
can be solved in $O((n-k)n\log n)$ time.
}

To solve {\bf MaxMin-Merge}$(k)$ more effective for small $k$, we consider
the problem directly and give an FPT algorithm\footnote{An FPT algorithm with respect to a parameter $k$ runs in time $O(f(k)n^c)$ 
for some computable function $f(k)$ and constant $c$, which can be very effective for small $k$.}.
Although FPT algorithms are intended for intractable problems, they are also useful for tractable problems, especially when they run in linear time for each fixed $k$.
To solve {\bf MaxMin-Merge}$(k)$, we use bounded search tree 
based on the following observation: 
{\em A minimum number must be merged with either its left or right neighbor}.

The above observation allows us to construct a search tree of height $k$ by finding a minimum number $x$ in the current string $S$, and recursively consider two cases with $k-1$ merges: merge $s$ with its left neighbor to obtain string $S_l$,
and merge $x$ with its right neighbor to obtain string $S_r$.
Recursively solve the problem on $S_l$ and $S_r$ for $k-1$, and return the better one of these two solutions as a solution for $S$.
Of course, we only consider $S_r$ (respectively $S_l$)  when $x$ is the first number (last number) of $S$.

Our search tree $T$ has $2^{k+1} - 1$ nodes, and the main cost of each node is to find a minimum number in the current string, 
which takes $O(n)$ time when we do it directly from $S$.
We now reduce the main cost of each node to $O(k)$ by using a binary heap to maintain potential minimum numbers of the current string.

As a merge changes the value of two numbers only, it is sufficient to consider a dynamic set of $2k$ smallest numbers in $S$, and we use a binary heap to maintain these numbers as candidates for a minimum number in the current string.
Also numbers of $S$ that are not left or right neighbors of these $2k$ smallest numbers will not be affected by optimal $k$ merges, and hence we may first reduce $S$ to a string with at most $4k$ numbers.
It takes $O(n)$ time to construct a binary heap for $S$, and $O(k \log n)$ time to find $2k$ smallest numbers in $S$ by using the heap.
Once we have found these $2k$ numbers, we construct a binary heap $H$ on them,
and use $H$ to dynamically find a minimum number in the current string.

For each node of the search tree $T$, it takes $O(\log k)$ time to obtain a minimum number $x$ of the current string, and $O(k)$ time to
update heap $H$ to heap $H_l$ for merging $x$  with its left neighbor
and heap $H_r$ for merging $x$ with its right neighbor.
Therefor each node of $T$ takes $O(k)$ time, and the whole algorithm takes $O(n + k\log n + 2^kk)$ time.
This indicates that for small values of $k$ (roughly speaking, $k \le \log_2n$),
our FPT algorithm performs better than our dynamic programming approach.

\thm{fpt}{
{\bf MaxMin-Merge}$(k)$ can be solved in $O(n + k\log n +2^kk)$ time,
which is linear time for each fixed $k$.
}

\section{Concluding remarks}

We have obtained efficient algorithms for {\bf Fewest-Merge}($b$) and 
{\bf MaxMin-Merge}($k$) that deal with merging adjacent numbers in a string.
Our algorithms explore connections of our problems with those concerning
partition of a string into pieces 
and selection of points from a set of points on a line.
We now conclude the paper with a few open problems.

\noindent {\bf Problem 1}. Is there a linear-time algorithm for {\bf Fewest-Merge}($b$)
on a cyclic string?

We can solve the problem in $O(n^2)$ time by considering $n$ possible ways to cut 
the cycle into a path and then use our $O(n)$-time algorithm for strings.
However, the trick for finding a substring in a cyclic string does not work here.

\noindent {\bf Problem 2}. In connection with {\bf MaxMin-Merge}$(n-k)$,
can we solve {\sc Cut String}$[k]$ in linear time, or just linear time for each fixed $k$?
And how about the problem on a cyclic string?

Here we are looking for an FPT algorithm of linear time for each fixed $k$ (uniformly linear-time algorithm).

\noindent {\bf Problem 3}. For {\bf MaxMin-Merge}$(k)$, is there a linear-time 2-approximation algorithm?

Although faster algorithms may be possible for the problem, a linear-time algorithm seems hard to reach, if not impossible. By lowering our demand to approximation algorithms, we should be able to obtain faster and simpler algorithms.

\section*{Acknowledgement}
The author thanks Emeritus Professor CAI Leizhen of the Chinese University of Hong Kong for his guidance in carrying out this summer project.

\nocite{*}

\printbibliography{}
\end{document}